\documentclass[aps,prl,preprint,groupedaddress]{revtex4}
\usepackage{graphicx}
\begin{document}

%\preprint{}


\title{Distribution of spectral-flow gaps in the Rashba model\\
with disorder: a new universality}

% Please use the appropriate macro for each type of information


\author{Daijiro Tobe$^{1}$, Mahito Kohmoto$^{1}$, Masatoshi Sato$^{1}$
and Yong-Shi Wu$^{1,2}$}
%\thanks{}
\affiliation{$^1$ Institute for Solid State Physics, University of Tokyo\\
5-1-5 Kashiwanoha, Kashiwa, Chiba 277-8581, Japan}
\affiliation{$^2$ Department of Physics, University of Utah, Salt
Lake City, UT 84112-0830, U.S.A.}


\date{\today}

\begin{abstract}
We report a study of disordered electron systems
with spin-orbit coupling on a cylinder using methods of
random matrix ensembles. With a threading flux turned on, the single
particle levels will generally avoid, rather than cross, each
other. Our numerical study of the level-avoiding gaps in the
disordered Rashba model demonstrates that the normalized gap
distribution is of a universal form, independent of the random
strength and the system size. For small gaps it exhibits a linear
behavior, while for large gaps it decays exponentially. A
framework based on matrix mechanical models is suggested, and is
verified to reproduce the universal linear behavior at small gaps.
Thus we propose to use the distribution of the spectral-flow gaps
associated with flux insertion as a new way to characterize 2d
random systems with spin-orbit coupling. The relevance and qualitative
implications for spin (Hall) transport are also addressed.
\end{abstract}

%\keywords{}


\maketitle


%\subsection{}
%\subsubsection{}

% If in two-column mode, this environment will change to single-column
% format so that long equations can be displayed. Use
%\begin{widetext}
%\end{widetext}

% figures should be put into the text as floats.
% Use the graphics or graphicx packages (distributed with LaTeX2e)
% and the \includegraphics macro defined in those packages.
% See the LaTeX Graphics Companion by Michel Goosens, Sebastian Rahtz,
% entire page. There is no need to do explicit centering.

{\it Introduction} In recent years, there is a surge of interests in
doing spintronics, i.e. to manipulate and to control electron spins
in solid state devices, with external electric, instead of magnetic,
fields. Potential advantages include efficient generation of spin
polarization inside the devices, handy manipulation of individual
spins at nano-scales, and desired reduction of heat dissipation
during information processing. (The last point is because, in
particular, the electric field can induce a transverse spin Hall
current, respecting time reversal invariance and therefore being
dissipationless \cite{MNZh03}.) Since there is no fundamental direct
coupling between spin and the electric field, the spin-orbit
coupling (SOC) plays a central role in inducing the spin transport.
For these reasons, the study of spin transport and, in particular,
the spin Hall effect in systems with SOC has recently attracted
much attention (for recent reviews, see e.g.\cite{Rashba06,ERH06}). In
the literature, a mechanism
for the spin transport due to SOC is called {\it extrinsic} or {\it
intrinsic}, depending on whether disorder (such as impurities or
imperfections) is involved or not. However, in a real sample there
is always disorder; to study possible interplay between the
intrinsic and extrinsic mechanisms, it is better to study them
simultaneously in one unified framework, rather than to study them
separately and add their contributions. In this paper we report such
a study of the systems with SOC using the methods of random matrix ensembles.

A prototype of the intrinsic SOC is the Rashba coupling in a two
dimensional electron gas. The intrinsic contribution \cite{SCNSJM}
to the spin Hall conductivity in this model, which has a universal
value $e/8\pi$, was found to be exactly cancelled by the extrinsic vertex
contribution from impurities in a perturbative approach
\cite{Mishch04,IBM04}. Naturally one asks whether the cancellation
persists beyond perturbation theory, and if so whether this points
to a new universality to be discovered in the disordered systems
with SOC concerning the spin (Hall) transport. These are the
questions we want to address.

We put the Rashba model with on-site randomness on a cylinder, and
mimic an infinitesimal electric field by turning on slowly a
magnetic flux threading the cylinder. The latter is equivalent to
imposing a twisted phase in the boundary condition. 
The same technique has been used before in
studying charge transport in the presence of localization
\cite{Thouless74}, as well as the integer \cite{Laughlin} and
fractional \cite{NTW85} quantum (charge) Hall effect. What we are
concerned about here is the behavior of level-avoiding gaps due to
disorder as the threading flux varies. (The relevance of the
level-avoiding gaps to the transport has been discussed before in
Refs. \cite{Thouless89} for the charge Hall effect and in Refs.
\cite{SSWH05,QWZh05,CWS05} for the spin Hall effect in this and
similar models.) We have carried out a numerical study and indeed
found a new {\it universality} in the random system with SOC. More
concretely, the distribution of the (normalized)
level-avoiding gaps is found to have a universal form, independent
of both the random strength and the system size. It is linear at
small gaps and decays exponentially at large gaps; and the average
gap for un-normalized distribution seems to tend to a finite value
when the system size becomes very large. In this paper we will
suggest a matrix mechanical model for the underlying random
ensemble, which is verified to exhibit the desired small-gap
behavior. Usually one uses the level-spacing statistics as a
characteristic of the universality in a random system or ensemble
\cite{Mehta}. Here our results suggest a different type of
characterization of random ensembles using the gap statistics in the
spectral flow due to flux threading (or twisted boundary
conditions).

{\it The Disordered Rashba Model} Consider the disordered Rashba
model \cite{Ando,SSWH05} on a square lattice with size $L\times L$:
\begin{eqnarray}\label{rashba}
H=-t\sum_{\langle i j \rangle}c_i^{\dagger}c_j +\sum_i w_i
c_i^{\dagger}c_i +V_{\rm SO} \left(i\sum_i c_i^{\dagger}\sigma_x
c_{i+\hat{y}} -i\sum_i c_i^{\dagger}\sigma_yc_{i+\hat{x}} +{\rm
h.c.}\right) ,
\end{eqnarray}
where $c_i=(c_{i\uparrow},c_{i\downarrow})$ are electron
annihilation operators for up-spin and down-spin, respectively,
$t$ the nearest-neighbor hopping, and $V_{\rm SO}$ the Rashba
spin-orbit coupling. Disorder is represented by uniformly
distributed on-site random potential $w_i \in [-W/2,W/2]$. We
impose the free boundary conditions in $x$-direction and the
twisted boundary conditions in $y$-direction:
$\Psi(i,j+L)=e^{i\theta}\Psi(i,j)$, so that the model is put on a
cylinder, which is threaded by a magnetic flux
$\Phi=\hbar\theta/e$.

We diagonalize the Hamiltonian numerically. The basic parameters
are $L$, $W$ and $V_{\rm SO}$. We have taken their ranges to be
$L$=12, 14, 16, 18, 20, 24, 28, 32, 36, 40, 48 and $W=0.05t, 0.1t, 0.2t,
0.4t, 0.6t, 0.8t, 1.0t$,
with the SOC fixed to be $V_{\rm SO}=0.5t$ unless
stated otherwise. We can safely assume that our system is in the
metallic phase, as the range of the randomness, $W$, is much less
than the critical value ($W_{\rm c}=5.875t$ for $V_{\rm
SO}=0.577t$) \cite{Ando} for the insulating phase.
\begin{figure}
\includegraphics[width=6cm]{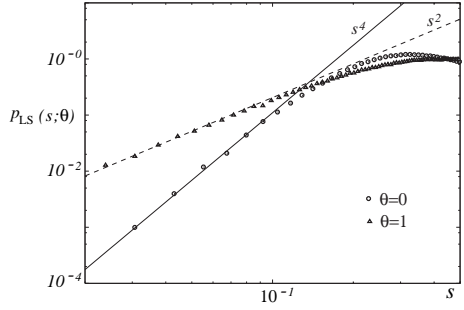}
\caption{\label{fig:ls_th_0}The level spacing distributions
$p_{\rm LS}(s;\theta)$ for $\theta=0$ and $\theta=1$. $L=24$ and
$W=0.2t$. For both data, the numbers of the level spacings
we take are about $10^6$.}
\end{figure}
We have first calculated the level spacing distribution,
$p_{\rm LS}(s;\theta)$ with $\theta$ fixed, where $s$ is the
spacing between adjacent levels. Indeed we have seen a flux
driven crossover from the symplectic to the unitary
ensemble. Some typical results are shown in
Fig. \ref{fig:ls_th_0} for $\theta=0$ and $\theta=1$. We find
the following power law behavior in the small-$s$ region,
\begin{eqnarray}
p_{\rm LS}(s;\theta)\propto
\left\{
\begin{array}{ll}
s^4 & \mbox{for $\theta=0, \quad {\rm symplectic}$}, \\
s^2 & \mbox{for $\theta=1, \quad {\rm unitary}$},
\end{array}
\right.
\end{eqnarray}
in agreement with the random matrix ensembles with the
corresponding symmetry \cite{Mehta}: When $\theta=0$,
the system has time-reversal symmetry, leading to
Kramer's degeneracy, which is broken when $\theta\neq 0$.

\begin{figure}
\includegraphics[width=13cm]{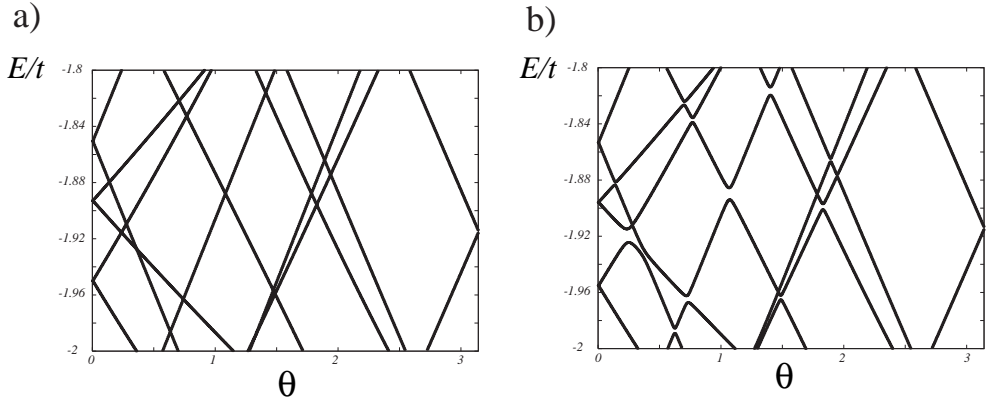}
\caption{\label{fig:level_diagram} Level diagrams for $L=10$
 a) without disorder ($W=0$) and b) with disorder ($W=0.1 t$).}
\end{figure}

{\it Spectral-Flow Gap Distribution} We are mainly interested
in the spectral flow due to a varying flux. Typical level-flux
diagrams are shown in Fig. \ref{fig:level_diagram}.
Without disorder (i.e. $W=0$), the energy levels cross each other
as $\theta$ is varied (Fig. \ref{fig:level_diagram}a), because
momentum $k_y$ is a good quantum number. With $W \neq 0$, however,
the translation symmetry in $y$-direction is broken by disorder,
and generically gaps open near the would-be level-crossings (Fig.
\ref{fig:level_diagram}b). To characterize the spectral flow, one
might try to extract from numerical data the statistical
distribution of the level-avoiding gap $g$ which is, however,
expected to depend on the parameters of the model. So a better way
to detect whether there is universality is to plot the
distribution $p_{\rm G}(\xi)$ of the {\it normalized} gap $\xi$,
the value of $g$ normalized by the average $\langle g \rangle$:
$\xi=g/{\langle g \rangle}$ with $L$ and $W$ fixed.

\begin{figure}
\includegraphics[width=13cm]{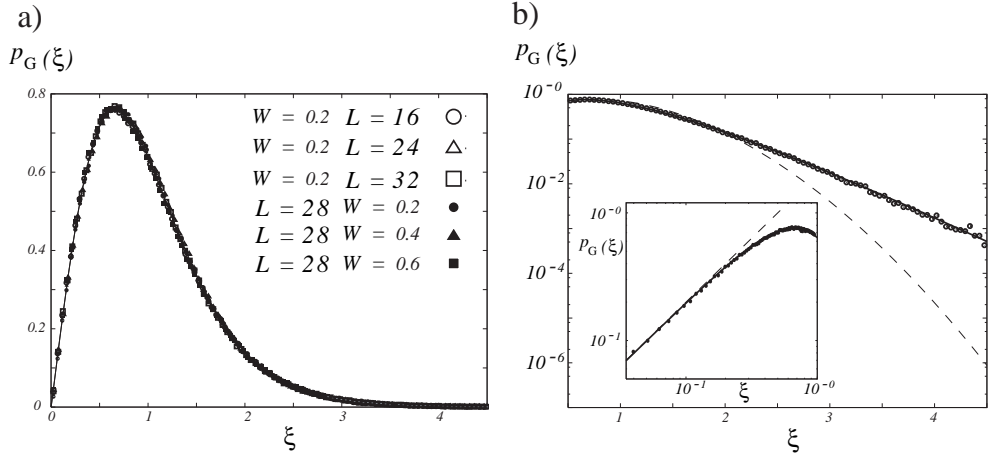}
\caption{\label{fig:level_avoiding_gap}a) The gap distributions
$p_{\rm G}(\xi)$ for various $L$ and $W$ (in units of $t$). The
solid curve is the function (\ref{distrib}).  
b) The gap distribution
$p_{\rm G}(\xi)$ for $L=24$ and $W=0.2t$. For comparison, the function
 (\ref{distrib}) (solid line) and the Gaussian unitary ensemble (GUE)
 surmise $(\pi \xi/2) e^{-\pi\xi^2/4}$ \cite{ZDK}(dashed line) are also
 shown.
The inset shows the small-$\xi$ behavior which can be fitted by a
linear behavior $\propto \xi$ (dashed line).
The number of gaps counted
 for each data point is greater than $5\times 10^6$.}
\end{figure}

Our numerical data for a wide range of $L$ and $W$, as exemplified
in Fig. \ref{fig:level_avoiding_gap}a, demonstrates a remarkable universal
behavior, namely the distribution, $p_{\rm G}(\xi)$, of the
normalized gaps neither depends on $L$ nor $W$. Indeed the
distribution is fit nicely by the following function:
\begin{eqnarray}
p_{\rm G}(\xi)=\frac{a \xi^2}{{\rm sinh}(b\xi)}, \quad {\rm with}
\quad a=\frac{2b^3}{7\zeta(3)}, \;\;  b=\frac{\pi^4}{28\zeta(3)},
\label{distrib}
\end{eqnarray}
where $\zeta(z)$ is the zeta function, and constants $a$ and $b$ are fixed by
the normalization conditions $\int_0^{\infty}d\xi p_{\rm G}(\xi)=1$ and
$\int_0^{\infty}d\xi \xi p_{\rm G}(\xi)=1$.
In Fig.\ref{fig:level_avoiding_gap}b, we show the data for the behavior
of the distribution at large and small gaps: It is linear at small gaps
and exponentially decays at large gaps.

\begin{figure}
\includegraphics[width=13cm]{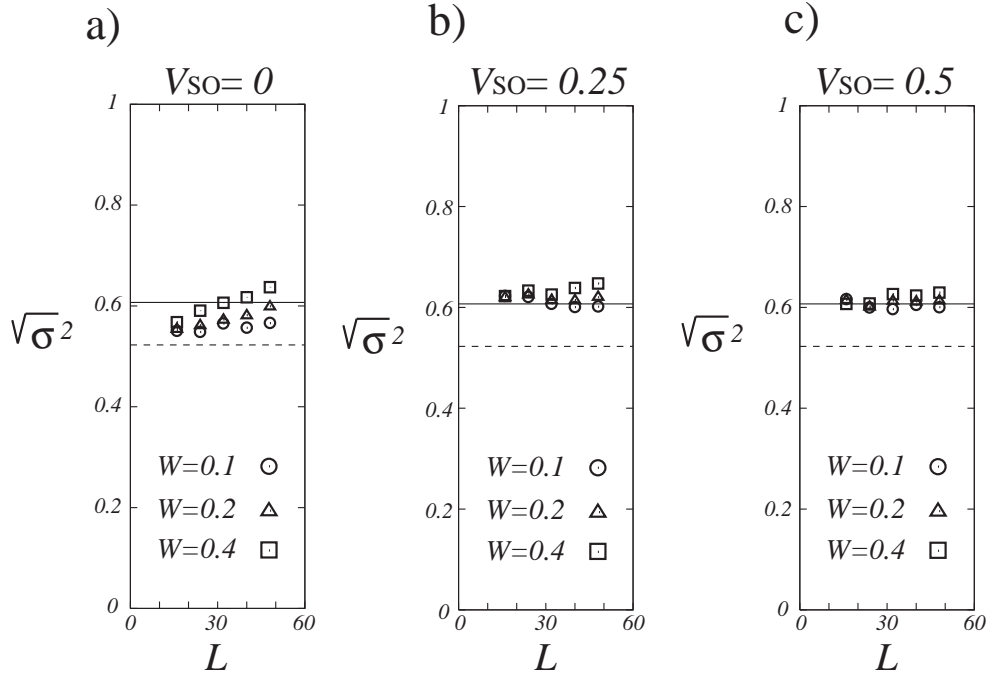}
\caption{\label{fig:withoutSO} The variances $\sqrt{\sigma^2}$ of the
 gap distribution a) without and b, c) with SOC. For comparison the
 variance of the
 function (\ref{distrib})(solid line) and that of the GUE
 surmise (dashed line) are also shown. }
\end{figure}

To examine whether the SOC is crucial for the appearance
of the universal behavior or not, 
we present in Fig.\ref{fig:withoutSO} some data for the variance,
$\sqrt{\sigma^2}=\sqrt{\int_0^{\infty}d\xi\xi^2p_{\rm G}(\xi)}$,
of the normalized gap distribution with and without the SOC,
respectively. While the data with $V_{\rm SO}=0.5t$ and $V_{\rm
SO}=0.25t$
support a universal behavior, those with $V_{SO}=0$ clearly show
non-universal $L$- and $W$-dependence.

\begin{figure}
\includegraphics[width=13cm]{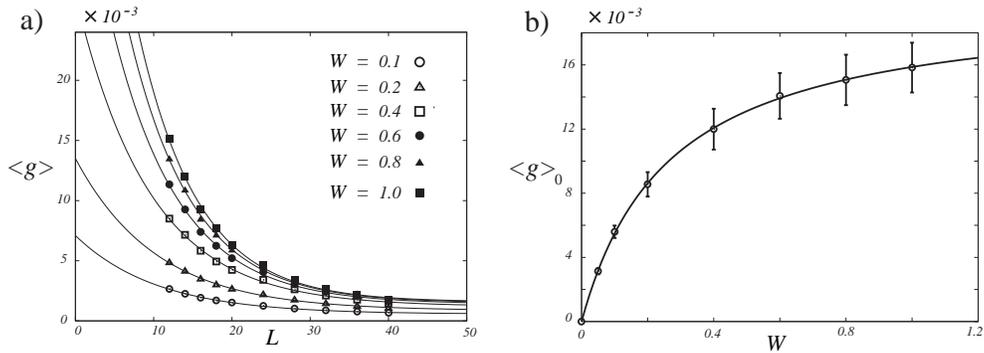}
\caption{\label{fig:av_g}a) Non-universal behaviors of the average
gaps $\langle g \rangle$. b) The $W$-dependence of the average gap
in the limit $L\rightarrow\infty$.}
\end{figure}

Numerically we have also checked the large-$L$ behavior of the
average gap $\langle g \rangle$, which is known to be
non-universal. Figure \ref{fig:av_g}a shows that it tends to a
non-zero value as $L\to \infty$ with $W$ fixed. This fact is
essential for the distribution of the normalized gap (\ref{distrib})
to make sense in the thermodynamic limit. When combined with the
Laughlin's argument, it has a profound implication concerning spin
Hall transport, which we are going to discuss below. 
In Fig.\ref{fig:av_g}b, we show the $W$-dependence of the limiting value
$\langle g\rangle_0$; the solid curve is the following function,
$\langle g\rangle_{0} =\alpha t \tanh\left(\beta \ln (\gamma W/t+1)\right)$
with $\alpha=2007.8$, $\beta=0.50$ and $\gamma=7.50$.

\begin{figure}
\includegraphics[width=8cm]{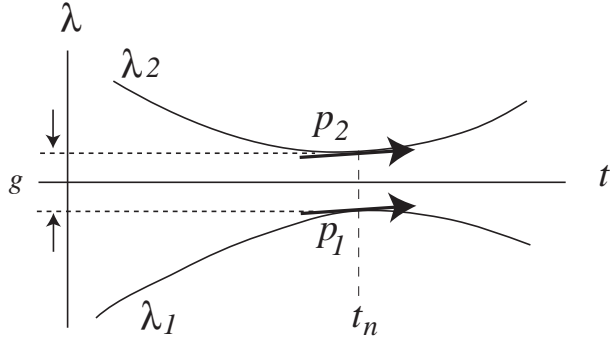}
\caption{\label{fig:avoid_crossing} An avoiding gap between $\lambda_1$
 and $\lambda_2$.}
\end{figure}

{\it New Universality of Random Ensembles} The above-discovered
universality of the distribution $p_{\rm G}(\xi)$ inspires us to
search for the random ensemble underlying it. We notice that the
spectral flow gaps we focus here are similar to the
so-called avoided crossings that have been extensively studied in
quantum chaos (see, e.g.\cite{Haake}). Our model contains a periodic (flux)
parameter $\theta$ in the twisted boundary condition. Since time
reversal invariance is respected only for $\theta=0$ and $\pi$, our system
should correspond to a random ensemble with unitary symmetry except
at $\theta=0$ and $\pi$, near which there are crossovers to symplectic
symmetry. The distribution of avoided crossings for Gaussian unitary
ensembles (GUE) was studied before in the literature of quantum
chaos \cite{Haake}. In Fig.\ref{fig:level_avoiding_gap}b, we
show the comparison of our data with the GUE surmise given in
Ref. \cite{ZDK}. Though the linear behavior in the small-$\xi$ region
agrees with the GUE studied in Refs. \cite{Wilkinson,AW}, the
exponential decay in the large-$\xi$ region certainly does not
\cite{ZDK}. Since the spectral flow is due to a varying parameter
$\theta$, a more natural setting should be a matrix quantum mechanical
model \cite{SLA}, in which the time variable $t$ may be identified with
$\theta$. Below let us show
that in a wide class of matrix mechanical models, the distribution
$p_{\rm G}(\xi)$ is indeed linear at small gaps and is independent
of either the size of matrices or the potential in the model. The
Lagrangian of the models we consider is
\begin{eqnarray}
L=\frac{1}{2}{\rm Tr}\dot{\Phi}(t)^2+{\rm Tr}V(\Phi(t))\ ,
\end{eqnarray}
where $\Phi(t)$ is an $N\times N$ {\it Hermitian} matrix, and ${\rm
Tr}V(\Phi)$ a potential invariant under the automorphism
$\Phi\rightarrow G\Phi G^{\dagger}$ with $G$ unitary. Writing
$\Phi(t)=U(t)^{\dagger}{\rm diag}(\lambda_1(t),\cdots
\lambda_N(t))U(t)$ in terms of its eigenvalues $\lambda_i(t)$ and
the unitary matrix $U(t)$ that diagonalizes it, the Lagrangian is
recast into
\begin{eqnarray}
L=\sum_i \frac{1}{2}\dot{\lambda}^2_i
+\sum_{i<j}\left|\lambda_i-\lambda_j\right|^2\omega_{ij}\omega_{ji}
+\sum_{i}V(\lambda_i)
\end{eqnarray}
with $\omega=iU\dot{U}^{\dagger}$. The corresponding Hamiltonian is
\begin{eqnarray}
H=\frac{1}{2}\sum_i p_i^2+\sum_i V(\lambda_i)
+\sum_{i<j}\frac{L_{ij}L_{ji}}{|\lambda_i-\lambda_j|^2},
\end{eqnarray}
where $p_i$ and $L_{ij}$ are the conjugate momenta of $\lambda_i$ and
$\omega_{ji}$, respectively.
The statistical properties are determined by the Gibbs measure,
\begin{eqnarray}
dG_N=\frac{1}{{\cal Z}_N}e^{-\beta H}\prod_i d\lambda_idp_i
\prod_{i<j}dL_{ij}dL_{ij}^{*},
\end{eqnarray}
where ${\cal Z}_N$ is the normalization constant. The un-normalized
gap distribution $n(g)$ is given by
\begin{eqnarray}
n(g)=N(N-1)\prod_{k=3}^{N}\int_{g/2<|\lambda_k|}dG_N\sum_{n}
\delta(t-t_n)\delta(-g/2-\lambda_1)\delta(g/2-\lambda_2),
\end{eqnarray}
where $t_n$ denotes a level avoiding point between $\lambda_1$ and
$\lambda_2$.
See Fig.\ref{fig:avoid_crossing}.
From the normalization condition, the gap distribution is
obtained by
\begin{eqnarray}
p_{\rm G}(\xi)=\langle g \rangle n(g)\left(
\int dg n(g)\right)^{-1}.
\end{eqnarray}
At the level avoiding points $t=t_n$, we have $p_1=p_2$ with
$\dot{p_2}-\dot{p_1}>0$.
Thus $\sum _n \delta(t-t_n)$ in the integral can be replaced with
\begin{eqnarray}
\sum_{n}\delta(t-t_n)&=&\delta(p_2-p_1)(\dot{p}_2-\dot{p}_1
+|\dot{p}_2-\dot{p}_1|)/2
\nonumber\\
&=&-\delta(p_2-p_1)\left(\frac{\partial H}{\partial(\lambda_2-\lambda_1)}
-\left|\frac{\partial H}{\partial(\lambda_2-\lambda_1)}
\right|\right),
\end{eqnarray}
where we have used the equations of motion for $p_1$ and $p_2$.
While the resulting integral is hard to perform, the small-$g$
behavior can be easily seen to be
\begin{eqnarray}
n(g)=A g,
\end{eqnarray}
where $A$ is a constant given by
\begin{eqnarray}
&&A=\frac{4N(N-1)}{\beta{\cal Z}_N}
\prod_{i=1}^{N}\int dp_i
\prod_{i<j}\int d\tilde{L}_{ij}\int d\tilde{L}_{ij}^{*}
\prod_{k=3}^{N}\int d\lambda_k\prod_{2<k<l}(\lambda_k-\lambda_l)^2
\prod_{k=3}^N\lambda_k^4
\nonumber\\
&&\times
\delta(p_2-p_1)
\exp\left[
-\beta \left(\sum_{i=1}^N \frac{p_i^2}{2}
-\sum_{i<j}\tilde{L}_{ij}\tilde{L}_{ji}-\sum_{k=3}^N V(\lambda_k)-2V(0)\right)
\right].
\end{eqnarray}
Thus we have a universal linear behavior for $p_{\rm G}(\xi)$ in the
small-$\xi$ region for any $N\ge 2$.
The insensibility of the linear behavior to the number of levels $N$
coincides with that to $L$ for the Rashba model as the number of levels
of the Rashba model is given by $2L^2$.

We note that the large-$\xi$ behavior of the random matrix quantum
mechanics does depend on the form of the potential $V(\lambda)$.
This is well-known for usual random matrix ensembles. For
example, for polynomial potentials, the kernel of the unitary
ensembles in the large $N$ limit has a universal form
corresponding to the Gaussian ensemble\cite{BZ}:
\begin{eqnarray}
K(x,y)=\frac{\sin \pi(x-y)}{\pi(x-y)};
\end{eqnarray}
but for logarithmically growing potentials, $V(\lambda)\sim
\frac{1}{2a}(\ln \lambda)^2 \quad (\lambda \gg 1)$, it shows
another universality \cite{MCIN}:
\begin{eqnarray}
K(x,y)=\frac{a}{\pi}\frac{\sin \pi(x-y)}{\sinh a(x-y)}.
\end{eqnarray}
It is plausible that there exists a class of potentials in matrix
mechanical models, which reproduce the spectral-flow gap
statistics we have found from our numerical calculations.

{\it Spectral-Flow Gaps and the Hall Transport} In the above we have
studied the distribution of spectral-flow gaps in a disordered
system with SOC, as a response to twisted boundary
conditions. This is in line with the Laughlin gauge argument
\cite{Laughlin} for the charge Hall effect, which has been recently
adapted to the case of the spin Hall effect \cite{SSWH05,QWZh05,CWS05}
for systems with SOC. Our data show no level crossing in the spectral flow, in
agreement with Refs. \cite{SSWH05,CWS05} for the disordered models.
Because of level avoiding, each level always flows back to its
original energy after the flux adiabatically increases from zero to
the unit flux quantum. In contrast to the quantum (charge) Hall
effect, in the present case the system is in the metallic phase, and
there are levels going across the Fermi level. At absolute zero
temperature and in the adiabatic limit, the levels that are below the
Fermi level
and never go across it will maintain its own contribution to the
spin Hall conductivity, similar to the case of the charge Hall
effect. The sum of these levels to the spin Hall conductivity of the
system is expected generically to have zero expectation value if the
position of the Fermi level is set randomly. However, during the
adiabatic variation of the flux, the instantaneous spin Hall
conductivity will fluctuate because of the levels that move out of
or dive into the Fermi surface. So at least for the cylindrical
geometry, the time-averaged spin Hall conductivity should be very
small, compared to the clean intrinsic limit.

Moreover, a finite electric field will induce the transition across the
Fermi surface between levels with sufficiently small gaps
(Landau-Zener-like tunneling). 
Also at finite temperature, thermal
fluctuations will cause similar transitions between levels with gaps smaller
than the average thermal energy. Therefore, the fluctuations in spin
Hall conductivity may be significant if the temperature is not too
low and the applied electric field is not too small. It is no doubt
that the knowledge we obtained in this paper on the
distribution of level-avoiding gaps and on the magnitude of the
average gap will be useful for detailed analyses of the disorder
effects on the fluctuations in spin Hall conductivity.
Here we only point out some features inferred from our results.
First, the level avoiding gap
distribution is of a universal form for the normalized gap, which is
linear at small gaps. 
So we expect that there is also a universality in the fluctuations in
$\sigma^{\rm spin}_{\rm H}$ if properly scaled.
In other words, the fluctuations may have a one-parameter scaling law
by the average gap at sufficiently low temperature, where the
Landau-Zener tunneling is dominating.
Second, since the average gap is non-vanishing in the large-$L$ limit,
we expect that the fluctuations in $\sigma^{\rm spin}_{\rm H}$ will be
governed by an exponential factor involving the temperature or the electric 
field over the average gap. 

{\it Summary and Discussions}
In this paper, we examined the distributions of level avoiding gaps in the
Rashba model in the presence of disorder. 
We found that the normalized distribution is of a universal form,
independent of the random strength and the system size, while the
average value of the gaps is non-universal.

A natural expectation is that our universal distribution works for
other models with SOC.
(For example, a disordered two-dimensional electron gas with Dresselhaus SOC
\cite{MC05}, a generalized Rashba model \cite{NSSM05}, a graphene model
with SOC \cite{KM05}, and a two dimensional hole gas \cite{BZ05}.)
We note that this does not mean the spin Hall conductivity itself
is universal: 
It depends on the average value of the level-avoiding gaps, thus it could be
model dependent.
For example, if the average value becomes zero in the thermodynamic
limit, a net spin Hall effect can be observed. 
In addition, what we discuss in the spectral flow
argument is not the spin Hall current itself but the spin accumulation
after the unit flux insertion.  
Because the spin is not conserved in general,
these two quantities are not necessarily in accordance with each other. 
This also supports the expectation that systems with different spin Hall
conductivities may have the universal gap distribution we discovered.

In the absence of disorder the Rashba model has
integrability of the momentum conservation which is the origin of the
level crossings.
The integrability may explain a mixed behavior of our distribution
between the Poisson distribution and the GUE one.
Indeed, it is known that the gap
distributions for fully chaotic systems without integrability show no
mixed behavior and are in good agreement
with the GUE surmise \cite{ZDK}. 

To conclude, we remark that the new universality we discovered in this
paper may have consequences concerning quantum transport in
mesoscopic systems with SOC. It may also provide helpful insight
into the interplay between localization and SOC in the spin Hall
effect of disordered systems.

{\it Acknowledgments} 
The computation in this work has been done using the
facilities of the Supercomputer Center, Institute for Solid State
Physics, University of Tokyo. 
The discussions with E. Mishchenko, D.N. Sheng, Z.Y.Weng and M.Wilkinson are
acknowledged.
Y.S.Wu is supported in part by the U.S. NSF
through grant No. PHY-0407187. This work was begun in summer 2005,
when he visited the ISSP, University of Tokyo. He thanks the host
institution for the financial support and warm hospitality. 

\bibliography{rashba}

\end{document}